\documentclass[EPJST]{sn-jnl}% Math and Physical Sciences Numbered Reference Style 
\usepackage{graphicx}%
\usepackage{multirow}%
\usepackage{amsmath,amssymb,amsfonts}%
\usepackage{amsthm}%
\usepackage{mathrsfs}%
\usepackage[title]{appendix}%
\usepackage{xcolor}%
\usepackage{textcomp}%
\usepackage{manyfoot}%
\usepackage{booktabs}%
\usepackage{algorithm}%
\usepackage{algorithmicx}%
\usepackage{algpseudocode}%
\usepackage{listings}%
\begin{document}
	
	\title[Article Title]{Tidal dissipation and synchronization of the temperate exo-Earth LP 791-18d}
	
	
	\author*[1]{\fnm{Sylvio} \sur{Ferraz-Mello}} \email{sylvio@iag.usp.br}
	
	\author[1]{\fnm{Thayn\'a} \sur{Menezes Bechara}}
	%\equalcont{The junior author contributed with checks and some codings.}
	
	\author[1]{\fnm{Raphael} \sur{Alves-Silva}}
	
	\affil*[1]{\orgdiv{Institute of Astronomy, Geophysics and Atmospheric Sciences}, \orgname{University of S\~ao Paulo},  \city{S\~ao Paulo}, \postcode{05508-090}, \state{SP}, \country{Brasil}}
	
	
	\abstract{The creep tide theory is used to explore several aspects of the tidal evolution of the planetary system of the M-star LP 791-18 . We discuss the early synchronization of the exo-Earth LP 791-18d and show that the trapping of its rotation in a 3:2 spin-orbit resonance would only have been possible if its eccentricity were approximately 0.04 or larger. The planet is likely in synchronous rotation. The perturbations of the other planets in the system do not allow the complete damping of the orbital eccentricity, and the resulting mechanical energy balance indicates that the tidal energy dissipated inside the planet may flow through the planetary surface at approximately 1 ${\rm W/m}^2$. } 
	
	%%================================%%
	%% Sample for structured abstract %%
	%%================================%%
	%
	
	
	
	\maketitle	
	\section{Introduction}\label{intro}
	Planetary tides affect the evolution of close-in exoplanets in several ways. We used the creep-tide theory \cite{rheo2013, rheo2015} to discuss these perturbations in the case of an Earth-like planet moving in a close-in orbit. The creep tide theory is suitable for this study. It involves only one free parameter: the relaxation factor $\gamma$, which is inversely proportional to the dynamic viscosity. It is important to know how far we can go with a theory that only uses the laws of Physics without the adoption of ad hoc models. Here, we use an approximation of the creep-tide theory, taking into account the bodies' density distributions, and using suitable approximations for the moments of inertia and fluid Love numbers \cite{hybrid}. In the case of the stellar tides, the creep-tide theory is almost equivalent to the constant time-lag (CTL) version of Darwin's theory, but in the case of nongaseous planetary bodies, there is no similar theory among the usual ones.
	
	The chosen planet for this application of the creep-tide theory is the temperate exo-Earth LP 791-18d. This planet, discovered in 2020 by Peterson et al. \cite{LP791}, belongs to a system of three planets with periods between 0.9 (a super-Earth) and 5.0 days (a sub-Neptune) orbiting a late M dwarf star. The planet LP 791-18d, a temperate Earth, is the intermediary one with a period of 2.75 days and is very close to the inner border of the host star's habitable zone. Another reason for this choice is the proximity of the host star to us, which makes us believe that this planetary system is likely to be chosen as a target in future investigations with powerful space telescopes. Last but not least, the very extended study of the physics of this planet included in the discovery paper shows that this planet has characteristics that make important new studies of its evolution.
	
	One of the difficulties in studying this system concerns the low mass of the host star. LP 791-18 is a late M star. The availability of information on these stars is far lower than the information existing, for instance, for solar-type stars. Fortunately, some efforts have been made recently to close this gap. One of them directly affects the calculations presented here. We recall that the formulas given by Bouvier et al. \cite{Bouvier} that allow us to model the evolution of rotation of stars are valid only in the mass range from 0.5 to 1.1 M$_\odot$ and cannot be applied in the case of red dwarf hosts. However, a recent paper by Engle and Guinan \cite{Guinan} gives some simple models for the age-rotation relationship observed for stars in the range M2.5--M6.5. This allowed us to extend the range of existing models down to 0.05 M$_\odot$ (see the Appendix).
	
	This paper considers, in Sec. 2, the perturbation of the orbit of LP 791-18d due to the other planets, mainly the sub-Neptune LP 791-18c. The analytical modeling of these perturbations indicates that the eccentricity oscillates around the value $e=0.00143$, which agrees with the value $e=0.0015 \pm 0.00014$ found by Peterson et al. \cite{LP791}. 
	The tidal evolution of the system is discussed in Sec. 3 with emphasis on the energy released on the planet, enhanced by the existence of the forced eccentricity.
	\footnote{Atmospheric tides were not considered, as this planet is not expected to have a significant atmosphere given the high-energy irradiation it receives. The paradigm of this class of objects is TRAPPIST-1c and observations of this planet from the James Webb Space Telescope (JWST) were consistent with a thinner atmosphere or a bare rock planet \cite{Zieba}.} 
	Sections 4 and 5 discuss the rotational evolution of the planet, the early synchronization of the exo-Earth LP 791-18d, and an analysis showing that the trapping of the rotation in a 3:2 spin-orbit resonance would only have been possible if the eccentricity were apr. 0.04 or larger. Finally, it presents some preliminary results indicating the fast damping of the planetary obliquity, followed by a section that collects the main conclusions of the paper. The appendix considers with some details technical questions related to the dependence of the lag on the relaxation factor, and to the intrinsic variations in the stellar rotation. The parameters used in the examples are collected at the end of the paper. 
	
	\begin{table}[h]
		\caption{The planetary system LP 791-18 (cf. NASA Exoplanet Archive)}\label{tab0}
		\begin{tabular}{lccc}
			\toprule
			&  \quad Planet b \quad & \quad Planet d \quad  & \quad Planet c \quad \\
			\midrule
			Mass ($m_\oplus$)  & $-$ & 0.9 $\pm$ 0.5  & 7.1 $\pm$ 0.7  \\
			Orbital Period (days)  & 0.9480    &  2.7534  &  4.9899\\
			Eccentricity  & $-$ & 0.0015 $\pm$ 0.00014    &  0.00008 $\pm$ 0.00004 \\
			\botrule
		\end{tabular}
	\end{table}
	
	\section{The forced eccentricities}\label{sec:ecc}
	The orbital eccentricity of a close-in exoplanet continuously decreases due to the planetary tide. 
	One important feature of the system hosted by the star LP 791-18 is the presence of at least 3 planets whose mutual perturbations prevent the eccentricities from being damped to zero by the tidal interaction with the host star. The residual forced eccentricity of LP 791-18d is 0.0015 \cite{LP791}. This was determined through several N-body simulations. In this paper, we determine the residual forced eccentricity of the planets using an analytical approach. We use the same equations used to determine the forced equations in the Laplacian resonances (see \cite{DGS}) with the proviso of not using the simplifications that stem from the Laplacian relation $n_1-2n_2=n_2-2n_3$. The main contributions for the forced eccentricities of the planets have frequencies $n_1-2n_2$ and $n_2-2n_3$. We may in analogous ways, consider terms with other frequency combinations, but simple calculations show that they are much smaller than the first-order considered ones.  
	
	The differential equations for these perturbations are:
	\begin{eqnarray}
		\mathrm{i} \frac{d\zeta_1}{d t}&=&-F_{12} e^{\mathrm{i} u} \nonumber \\
		\mathrm{i} \frac{d\zeta_2}{d t}&=&-G_{12} e^{\mathrm{i} u} - F_{23} e^{\mathrm{i} u'} \label{eq:rhs}\\
		\mathrm{i} \frac{d\zeta_3}{d t}&=&- G_{23} e^{\mathrm{i} u'} \nonumber  
	\end{eqnarray}
	where $\zeta_i=e_i.e^{\mathrm{i} \varpi_i}$ $(i=1,2,3)$ are the complex composites of the osculating Keplerian variables related to the eccentricity and periastron, $\mathrm{i}=\sqrt{-1}$, $u=(n_1-2n_2)(t-t_0)$, and $u'=(n_2-2n_3)(t-t_0)$. The $n_i$ are the mean-motions and $F_{ij}$ and $G_{ij}$ are functions of the Laplace coefficients:
	\begin{eqnarray}
		F_{ij}&=-&\frac{1}{2} \frac{\mathcal{G}m_j}{n_ia_i^2a_j}\left(
		4b_{1/2}^2 + \alpha \frac{d b_{1/2}^2}{d \alpha} \right)\\
		G_{ij}&=& \frac{1}{2} \frac{\mathcal{G}m_i}{n_ja_j^3}\left(
		3b_{1/2}^1 + \alpha \frac{d b_{1/2}^1}{d \alpha} \right) -\frac{1}{2}\frac {\mathcal{G}m_i}{n_ja_ja_i^2}
	\end{eqnarray}
	where $a_i$ are the semi-major axes, $\alpha=a_i/a_j$\ $(j=i+1)$, $b_{1/2}^k(\alpha)$ are the Laplace coefficients, $m_i, m_j$ are the planet masses, and $\mathcal{G}$ is the gravitational constant \cite{DGS}.  
	
	\begin{figure}
		%	\begin{minipage}{\columnwidth}
			\centering
			%	\end{minipage}
		\includegraphics[width=6cm]{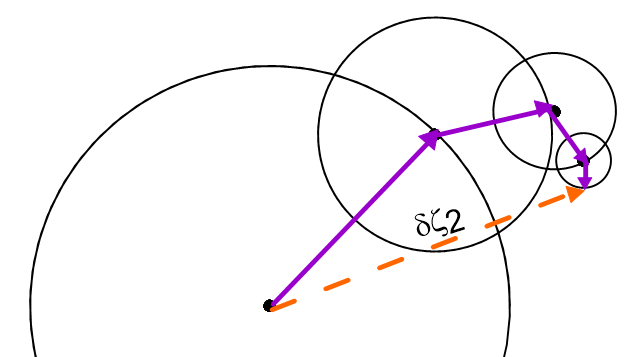}
		\caption{The osculating eccentricity is the modulus of a sum of complex functions.}
		\label{fig:plot2}       
	\end{figure}
	
	The homogeneous part of the differential equations is not written because we are not interested in the free eccentricities. Additionally, the corrections they introduce in calculating the forced eccentricities can be neglected because of the very small planetary masses.
	The integration is simple:
	\begin{eqnarray}
		\delta {\zeta_1}&=&\frac{F_{12}}{\mu} e^{\mathrm{i} u} \nonumber \\
		\delta {\zeta_2}&=&\frac{G_{12}}{\mu}e^{\mathrm{i} u} + \frac{F_{23}}{\mu'} e^{\mathrm{i} u'}\\
		\delta {\zeta_3}&=& \frac{F_{23}}{\mu'} e^{\mathrm{i} u'} \nonumber  
	\end{eqnarray}
	where $\mu=n_1-2n_2$ and $\mu'=n_2-2n_3$. 
	
	The actual calculations in the system LP 791-18 give the leading terms:
	\begin{eqnarray}
		\delta {\zeta_1}&=& 0.00019 e^{\mathrm{i} u} \nonumber \\
		\delta {\zeta_2}&=&-0.00143 e^{\mathrm{i} u} + 0.00013 e^{\mathrm{i} u'}\\
		\delta {\zeta_3}&=& 0.00084 e^{\mathrm{i} u'}. \nonumber  
	\end{eqnarray}
	
	In Laplacian systems, the (equal) divisors $\mu$, $\mu'$ are generally very small and strongly magnify these perturbations (as in the case of the Galilean satellites)\footnote{$\mu=2.063 {\rm d}^{-1}$ and $\mu'=-0.236 {\rm d}^{-1}$}. Here, they are not equal and not so small, but small enough to make these terms dominate over those issued from other frequency combinations in the low-order perturbations in eccentricity. 
	
	In general, if 
	\begin{equation}
		\zeta_j = A_0 e^{\mathrm{i} a_0 t} + \sum_{k\in N}
		A_k e^{\mathrm{i} a_k t}
	\end{equation}
	and $A_0 > \sum_k A_k$, then the average osculating eccentricity is $ \langle |\zeta_j| \rangle = A_0$. The other terms will only contribute periodic oscillations around the average (see Fig. \ref{fig:plot2}).
	
	We emphasize the agreement of the result for $\delta\zeta_2$ with the result obtained by Peterson et al. \cite{LP791} from N-body simulations (0.0015).
	
	\subsection{The free eccentricities}
	
	When homogeneous terms $\sum_j c_{ij}\zeta_j$ are considered in Eqn. \ref{eq:rhs} instead of the forced terms, we obtain a linear system whose eigenvectors, conveniently normalized, are the so-called free eccentricities (see \cite{DGS}). 
	
	\begin{figure}
		%	\begin{minipage}{\columnwidth}
			\centering
			%		\framebox[\columnwidth][c]{\raisebox{0pt}[20mm][20mm]{Ecc18d.eps}}
			%	\end{minipage}
		\includegraphics[width=8cm]{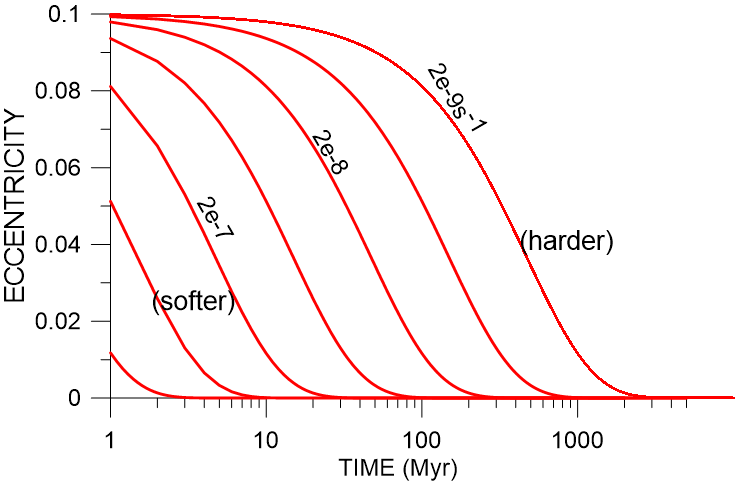}
		\caption{Damping of the eccentricity of LP 791-18d by the planetary tide for relaxation factors covering the range observed in the Solar System. Initial eccentricity: 0.1.}
		\label{fig:Ecc18d}       
	\end{figure}
	
	Fig. \ref{fig:Ecc18d} shows how a non-forced component of the eccentricity of planet LP 791-18d behaves under the perturbations of the planetary tide. The creep-tide theory was used to compute the evolution of the planet's eccentricity. The problem with this calculation is our ignorance of the relaxation factor of rocky exoplanets. We have calculated several cases with relaxation factors ranging from very low values ($2\times 10^{-9} {\rm s}^{-1}$), the relaxation factor of the Moon, to values slightly higher than the largest values determined for the planetary satellites of the Solar System (see \cite{rheo2013}). We recall that for solid Earth, $\gamma$ lies in the range $ 0.9 - 3.6 \times 10^{-7} {\rm s}^{-1}$.
	
	Figure \ref{fig:Ecc18d} favors a quick damping of any non-forced eccentricity that could have been excited by some event in the early ages of the system. However, we cannot exclude the possibility that the planet is hard enough to have a smaller relaxation factor, thus able to keep a non-forced eccentricity for a longer time. 
	
	The observations do not give any indication of a significant free eccentricity \cite{LP791}. Therefore, in the continuation of this paper, we assume that the only eccentricities remaining in the system are those forced by the mutual perturbations calculated in this section.

	\section{Tidal evolution of the system. Energy balance}
	
\begin{figure}
		%	\begin{minipage}{\columnwidth}
			\centering
			%		\framebox[\columnwidth][c]{\raisebox{0pt}[20mm][20mm]{Plot18d.eps}}
			%	\end{minipage}
		\includegraphics[width=8cm]{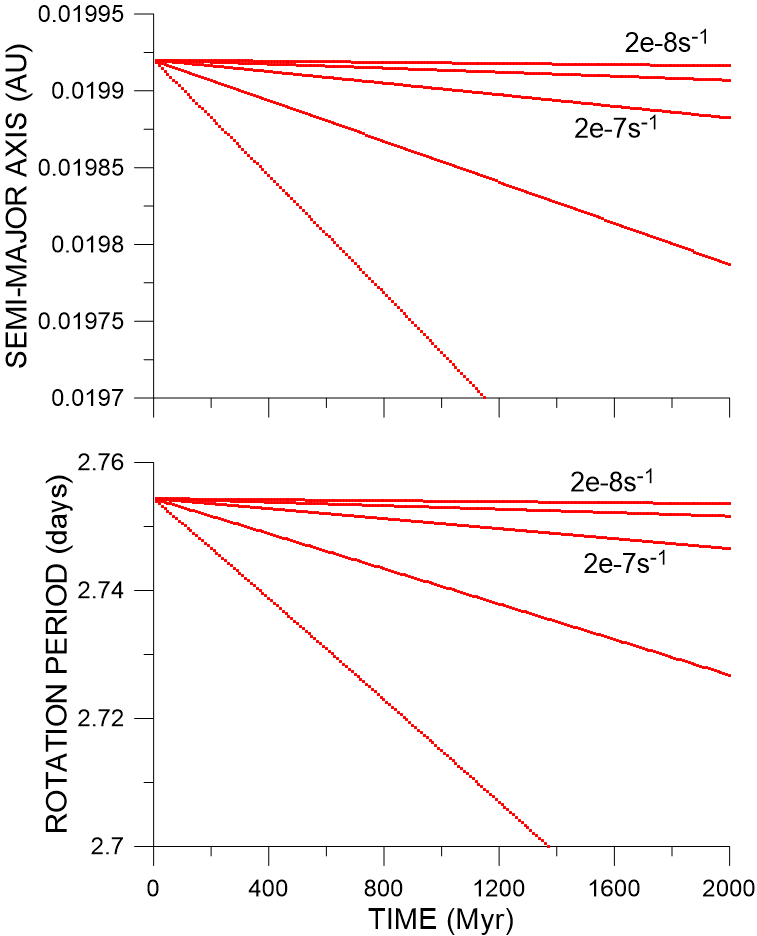}
		\caption{Tidal evolution of the semi-major axis and planetary rotation period of LP 791-18d for relaxation factors covering  the range  $ 2 \times 10^{-8} - 2 \times 10^{-6} {\rm s}^{-1}$ in the future. N.B. The initial orbital period is 2.7534 days.}
		\label{fig:Plot18d}       
	\end{figure}

	We may simulate the tidal evolution of the three-planet system at some time in the future. To avoid rapid damping of the eccentricities, which makes the tidal effects negligible, the equations are modified to allow the eccentricities to remain equal to the average forced ones as given in Section \ref{sec:ecc}. Variations of the semi-major axis and planetary rotational period are shown in Fig. \ref{fig:Plot18d}. We note that the variations in the planet's rotation closely follow the variations in the semi-major axis. This is so because the planet's rotation is trapped in a stationary state. In this case
	\begin{equation}
		{\Omega}_{\rm pl} \simeq n+6\frac{\gamma^2}{n}e^2
	\end{equation}
	where $n$ is the orbital mean motion of the planet.
	This state is almost synchronous. It is not exactly synchronous because the orbit is not circular. The forced eccentricity of 0.0015 does not allow for the full vanishing of the tidal torque and induces a small difference in the rotation speed. However, this would only happen if the planet were kept perfectly symmetrical all the time. The almost synchronization may entail the outcrop of an asymmetry in the planet's mass distribution able to counterbalance the small torque surplus due to the forced eccentricity. Therefore, the planet's rotation is expected to become perfectly synchronous with the orbital motion with, at most, a physical libration. A libration of this kind is observed on the Moon.   
	
	The stellar rotational period also shows a variation, but it is not due to tidal interactions and just comes from the angular momentum loss of the star due to stellar winds (see Appendix \ref{brake}). We emphasize that in these calculations, as well as in all other calculations done in this paper, the stellar and planetary tides are both considered. 
	
	\subsection{Energy dissipation}
	
	The main constituents of the system's mechanical energy are the orbital energy and the rotational energies. If we neglect the variations in the body geometry, the time derivatives in the energies are given by:
	\begin{equation} 
		\dot{\cal{W}}_{\rm orb}=\frac{\mathcal{G}Mm}{2a^2}\dot{a}, \qquad
		\dot{\cal{W}}_{\rm rot}= I_0\Omega\dot{\Omega}
		\label{eq:diss}
	\end{equation}
	($M$ is the star's mass, $\Omega$ is the rotational velocity, and $I_0$ is the axial moment of inertia of the considered body). Their variations lead to a monotonic decrease in the system's mechanical energy (see \cite{rheo2018}). They may be determined from the quantities $\dot{a}$ and $\dot\Omega$ obtained from the results shown in Fig. \ref{fig:Plot18d}. In the intermediate case corresponding to an Earth-like relaxation factor ($2\times 10^{-7} {\rm s}^{-1}$), we obtain
	$$\dot{\cal{W}}_{\rm orb} \sim  -5.5 \times 10^{14}{\rm W} $$
	$$\dot{\cal{W}}_{\rm rot\ pl} \sim +2.6 \times 10^{9} {\rm W}.$$
	The mechanical energy lost is liberated inside the planet and may flow through its surface at a rate $\sim 1.0 \ {\rm W/m}^2$. 
	
	In general, the energy lost by the orbit may contribute to the heat release in both the star and the planet. In this case, calculations have shown that the variation $\dot{a}$ is almost completely due to the planetary tide. It does not change when we neglect the stellar tide in the calculations. 
	
	This result is in good agreement with the results found by Peterson et al. \cite{LP791}: $0.2-0.9 \ {\rm W/m}^2$. The slight difference arises from the fact that the chosen relaxation factor does not precisely correspond to the quality factor Q and the tidal model adopted in \cite{LP791}.  With such dissipation in its interior, the planet is expected to present volcanism (as Io). The tidal heat must flow equally in both hemispheres and may have some influence on the temperature of the planet's dark side. 
	
	We may recall that the dissipated energy in synchronous rocky planets is more or less proportional to the relaxation factor. A softer planet would dissipate much more. However, at least in the Solar System, no rocky body is known with a relaxation factor larger than $10^{-6} {\rm s}^{-1}$. Not even Io, whose heat flow is evaluated as twice stronger than the value found here for LP 791-18d. In addition, we may mention that $\gamma = 2\times 10^{-7} {\rm s}^{-1}$ corresponds to a dynamic viscosity $6 \times 10^{17}\ {\rm Pa.s}$ which is within the range adopted by Léger et al. \cite{Corot7} for the viscosity of the super-Earth CoRoT-7b.      
	
	\section{Synchronization}
	
\begin{figure}[h]
		%	\begin{minipage}{\columnwidth}
			\centering
			%		\framebox[\columnwidth][c]{\raisebox{0pt}[20mm][20mm]{Lock05.eps}}
			%	\end{minipage}
		\includegraphics[width=8cm]{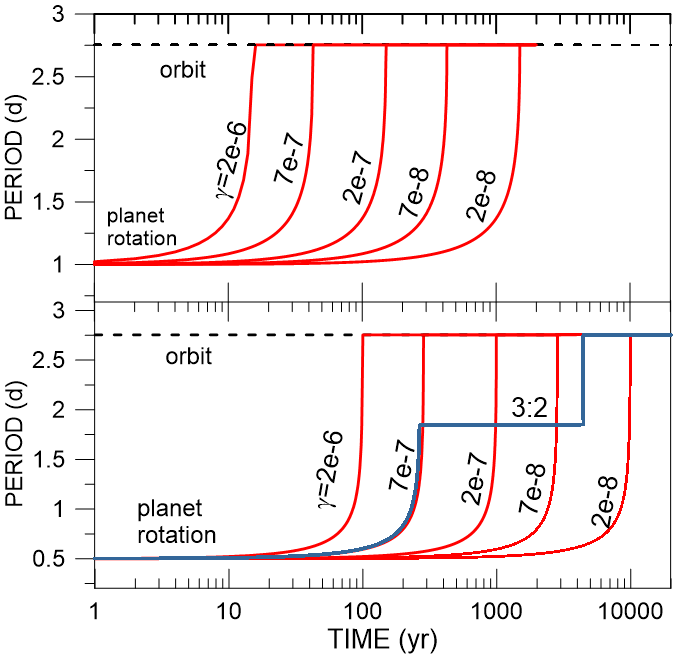}
		\caption{Tidal synchronization of the planet rotation. Initial rotation period: \textit{Top}:1.0 day \textit{Bottom}: 0.5 day. Initial eccentricity: e=0.0015 (red) and e=0.07 (blue). The labels indicate the planetary relaxation factors in the solutions shown.}
		\label{fig:Lock}       
	\end{figure}

	We collect in this Section the results obtained in various simulations showing the early synchronization of the planet LP 791-18d. It is not possible to define a synchronization time in an absolute way because it strongly depends on the adopted initial rotation. For this reason, we performed two different series of experiments with two different initial periods: 0.5 days and 1 day. 
	They are shown in Fig. \ref{fig:Lock}. In the two panels, the red lines show simulations in which the eccentricity was taken as equal to the forced value 0.0015. The adopted eccentricity does not influence the result except that with a higher eccentricity (blue line), the solution may remain trapped in an intermediary spin-orbit resonance for a long time (see Sect. \ref{sec:trap}).     
	
	The results shown in Fig. \ref{fig:Lock}\textit{Top} may be compared to those obtained by Barnes \cite{Rory} for the planet Proxima b, whose characteristics are very similar to LP 791-18d, but with a larger orbital period. The solutions constructed by Barnes with an initial period of 1 day may be reproduced with the theory used here with $\gamma = 3 - 5 \times 10^{-6} {\rm s}^{-1}$. This range of relaxation factors corresponds to time lags a little smaller than the 640 seconds adopted by Barnes. A more complete comparison is impaired by the fact that Barnes' solutions were constructed using the constant time lag (CTL) version of Darwin's theory, in which the phase lag $\varepsilon_0$ is proportional to the semi-diurnal frequency $\nu=2(\Omega-n)$. In the creep-tide theory, on the contrary, when $\gamma < < \nu$, the phase lag is rather inversely proportional to the semi-diurnal frequency (see Appendix \ref{Q}).
	
	\section{Spin-orbit resonance trapping}\label{sec:trap}
	
	Usual studies of the spin-orbit resonance consider an asymmetric companion (see \cite{Celletti}). The asymmetry, often characterized by the existence of a $J_{22}$-term, is responsible for a force that counterbalances the tidal forces, allowing the rotation of the body to be synchronized with the orbital motion (\cite{FRH, Calleg, Adrian}). The differential equations are second-order differential pendulum-like equations. However, if an asymmetry does not exist a priori, the dynamics is rather ruled by a first-order differential equation \cite{Correia, rheo2015} of the form
	\begin{equation}\label{eq:Omegadot}
		\langle \dot\Omega \rangle = -\mathcal{A} \sum_{k \in \mathbb{Z}} E_{2,k}^2\frac{\gamma(\nu+kn)}{\gamma^2+(\nu+kn)^2}
	\end{equation}
	\cite{rheo2013, hybrid} where $\mathcal{A}$ is a positive coefficient depending on the dynamical parameters of the system and $E_{2,k}$ are known functions of the orbital eccentricity (see Appendix D).
	
	\begin{figure}
		%	\begin{minipage}{\columnwidth}
			\centering
			%		\framebox[\columnwidth][c]{\raisebox{0pt}[20mm][20mm]{Omegadot1.eps}}
			%	\end{minipage}
		\includegraphics[width=10cm]{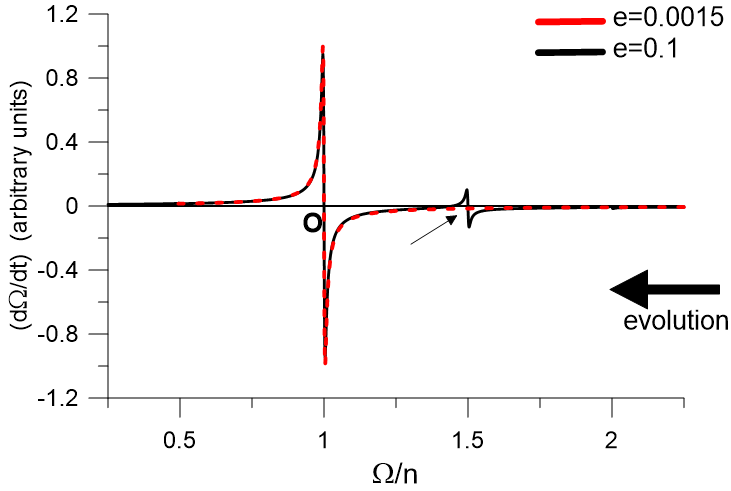}
		\caption{Phase plane  $(\Omega, \langle \dot\Omega \rangle)$ of Eqn. \ref{eq:Omegadot} in two cases: e=0.0015 (red) and e=0.1(black). In both cases $\gamma = 2\times 10^{-7}\ {\rm s}^{-1}$. The arrow shows the stable equilibrium corresponding to a 3/2 spin-orbit resonance.}
		\label{fig:Omegadot1}       
	\end{figure}
	
	Figure \ref{fig:Omegadot1} shows the phase plane of Eqn. \ref{eq:Omegadot} in two cases. In both cases, the evolution starts with a fast planet rotation, i.e., $\Omega > n$. In the case of low eccentricity (red curve), $\dot{\Omega} < 0$ for all values on the right of $\mathbb{O}$ and $\dot{\Omega} > 0$ on the left of $\mathbb{O}$, that is, $\mathbb{O}$ is a stable equilibrium solution. This is the synchronous solution discussed above. For completeness, we just remember that it is not exactly at $\Omega/n=1$ but slightly to the right of this point. When $e_0=0.1$ (black curve), an anomaly appears at $\Omega/n=3/2$. In that position, there appears a crest that crosses the axis $\dot{\Omega}=0$ and reaches positive values of $\dot{\Omega}$. This means that in this interval, one new stable equilibrium point appears. This point, indicated by an arrow in Fig. \ref{fig:Omegadot1}, is exactly the equilibrium point corresponding to a spin orbit resonance similar to Mercury's. 
	
	A consequence of the 3/2 crest can be seen in the blue line of Fig. \ref{fig:Lock}\textit{Top}. That line shows a simulation with $\gamma=2 \times 10^{-7}\ {\rm s}^{-1}$ with an initial period of 1 day and eccentricity $e_0=0.07$. Initially, this solution closely follows the solution for $e_0=0.0015$. The difference between the two is minimal. However, when the solution with $e_0=0.1$ reaches 2/3 of the planet's orbital period, it remains trapped in the corresponding spin-orbit resonance and remains there for a long time. Since the planetary tide also damps the orbital eccentricity of the orbit (see Fig. \ref{fig:Ecc18d}), the height of the crest is also continuously damped, and at a given point, the stable equilibrium disappears. $\Omega$ is again decreasing, and the solution evolves toward synchronization.

\begin{figure}[h!]
		%	\begin{minipage}{\columnwidth}
			\centering
			%		\framebox[\columnwidth][c]{\raisebox{0pt}[20mm][20mm]{zoom1.eps}}
			%	\end{minipage}
		\includegraphics[width=9cm]{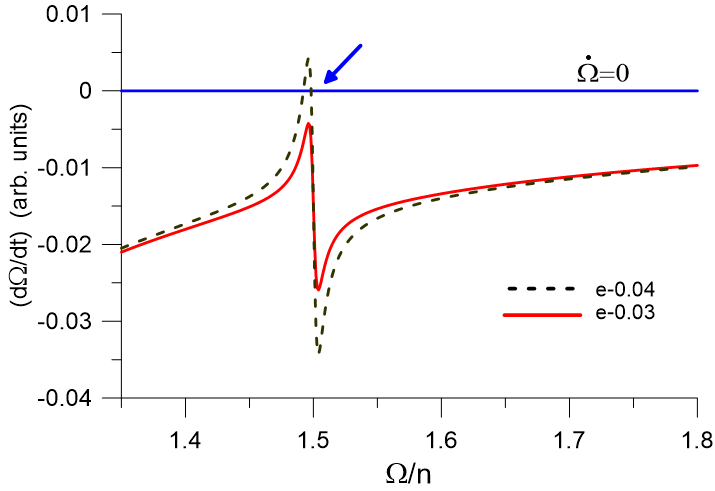}
		\caption{Phase plane  $(\Omega, \langle \dot\Omega \rangle)$ of Eqn. \ref{eq:Omegadot} in two cases: e=0.03 (red) and e=0.04(dashed black). In both cases $\gamma = 2\times 10^{-7}\ {\rm s}^{-1}$. The arrow shows the stable equilibrium corresponding to a 3/2 spin-orbit resonance.}
		\label{fig:zoom1}       
	\end{figure}

	The trapping-distrapping dynamics is better understood when looking at Fig. \ref{fig:zoom1}. In this figure, we show the plot of the curve $(\Omega, \langle \dot\Omega \rangle)$ in the phase plane of Eqn. \ref{eq:Omegadot}, in two cases: e=0.03 (red) and e=0.04 (black). In both cases $\gamma = 2\times 10^{-7}\ {\rm s}^{-1}$. Both curves show a crest at $\Omega/n=3/2$. However, in one case (e=0.03), the crest remains in the negative half-space all the time, while in the other (e=0.04), it crosses the axis $\dot{\Omega}=0$ and reaches positive values that allow the rising of a stable equilibrium solution.  
	
	Therefore, if the solution reaches the point $\Omega=3n/2$ with an eccentricity $e = 0.04$ or greater, the planet's rotation is trapped in the resonance.\footnote{Since, in first approximation, the energy dissipation is proportional to $e^2$ and the eccentricity necessary to have a 3/2 spin-orbit resonance is more than 20 times larger than the observed eccentricity, the dissipation in the resonant case would be some hundred times larger than it is in the current synchronous case.} If, subsequently, the eccentricity decreases and approaches 0.03, the solution escapes the spin-orbit resonance and evolves up to become synchronous.
	
	\begin{figure}[h!]
		%	\begin{minipage}{\columnwidth}
			\centering
			%		\framebox[\columnwidth][c]{\raisebox{0pt}[20mm][20mm]{zoomGamma.eps}}
			%	\end{minipage}
		\includegraphics[width=9cm]{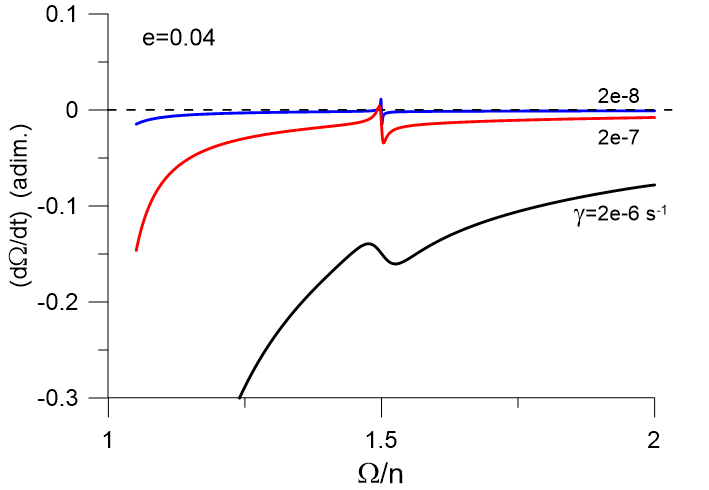}
		\caption{Dependence of the phase diagram  $(\Omega, \langle \dot\Omega \rangle)$ of Eqn. \ref{eq:Omegadot} on the relaxation factor $\gamma$. The labels indicate the planetary relaxation factors and the eccentricity of the cases plotted.}
		\label{fig:zoomGamma}       
	\end{figure}
	
	Finally, Fig. \ref{fig:zoomGamma} shows the influence of the planetary relaxation factor on the existence of the spin-orbit resonant solution. We see that the probability of being trapped in a spin-orbit resonance increases when the planet is harder (smaller relaxation factors). This figure considered three different cases with an eccentricity of 0.04. We see that when the planet is softer and $\gamma = 2\times 10^{-6}\ {\rm s}^{-1}$, the anomaly in the curve is striking but not enough to allow positive values of $\dot\Omega$ to be reached. In calculations with smaller eccentricities, the crest is no longer visible, and no trapping in spin-orbit resonance is possible. 
	
	These calculations and the fact that no significant eccentricity could be determined from the observations allowed us to say that the planet's rotation may be synchronized with the orbital motion.   
	
	\section{Planetary obliquity}
	
	All results shown above were derived using a coplanar theory in which the equator of the bodies and the orbit lie in the same plane. 
	The validity of this assumption was verified with an extension to the full 3D case recently published \cite{3D}. The 3D equations were used to study the variation in obliquity of planet LP 791-18d in cases with a small orbital inclination ($\sim 0.01$ rad) and two initial planetary obliquities: 30 and 45 degrees, respectively. 
	
	In both cases, the planetary obliquity is quickly damped to zero, reaching less than 0.05 degrees in about one thousand years. The stellar obliquity was taken close to 10 degrees. It just exhibited damped oscillations of small amplitude in this time interval.
	
	\section{Conclusion}
	
	An important conclusion is the synchronization of the planetary rotation. The trapping of the rotation in a different spin-orbit resonance would only have been possible if the eccentricity were $\sim 0.04$ or larger.
	On one hand, the damping of the orbital eccentricity is expected to bring it below the said limit in a time less than the age of the system,
	To be able to remain above that limit, the body should be much harder than assumed here (i.e., it should have a smaller $\gamma$). However, observations would have detected an eccentricity significantly higher than the forced eccentricity, and this was not the case \cite{LP791}. 
	
	The results found for LP 791-18d must reproduce themselves in other systems with similar characteristics. 
	All exo-Earths with very low eccentricity and close enough to the host star are expected to be synchronous. To remain trapped in a possible spin-orbit resonance, the orbit may have been excited to a significant eccentricity, and the planet must have a very low relaxation factor to avoid damping the eccentricity to low values. We note that the actual condition for remaining trapped in a spin-orbit resonance is that the osculating eccentricity has never been below its critical value. Indeed, once the system escapes one resonance, it can no longer be trapped again in the same resonance.   
	
	The tidal heat dissipation in LP 791-18d is strong and seems to be of the same order as that observed in Io. If the crust temperature is less than the solidus temperature and the conductivity is not large enough to allow efficient transport of the produced heat to the exterior, volcanism is expected to occur through volcanic craters or cracks in the crust. 
	
	The calculations supporting the above conclusion were done with a version of the creep tide theory allowing to take into account the density radial distribution in the bodies. However, it was constrained by the assumption of vanishing planetary obliquity. Calculations done with various examples with a full 3D theory indicate that an initial planetary obliquity would be damped in a very short time. However, the number of examples calculated and the timespan covered by these calculations do not allow for a more general statement. 
	
	The stellar tide was also considered, but given the small mass of LP 791-18d, it is negligible and does not significantly impact the system's evolution, even when the effects are enhanced by adopting a smaller stellar relaxation factor. 
	
	\begin{appendices}
		\section{Stellar rotation}\label{brake}
	Because of the small mass of the planet, the tide on the star almost does not contribute to the system's evolution. However, it was included in all calculations. The main component of the stellar tide is the semi-diurnal tide of frequency $\nu_*=2(\Omega_*-n)$. The calculations must consider the loss of stellar angular momentum due to the star's activity. Engle and Guinan \cite{Guinan} determined the relationship between the age and rotation period of red dwarf stars by analyzing their rotations and the ages of star groups with which they may be associated. In the range between M4 and M6.5, for stars with periods less than 175 days, they have found
		\begin{equation}
			\log t_{(Gyr)} = A P_{* (\rm days)} + B \label{eq:logtGyr}
		\end{equation}
		where
		\begin{eqnarray}
			A&=&0.0251 \pm 0.0022  \nonumber \\
			B&=&- 0.1615 \pm 0.0309    \nonumber
		\end{eqnarray}
		for $P_{\rm rot} < 25.4500 \pm 2.4552 $ days, and
		\begin{eqnarray}
			A&=&0.0039 \pm 0.0031  \nonumber \\
			B&=&0.3780 \pm 0.0535   \nonumber
		\end{eqnarray}
		for $P_{\rm rot} \ge 25.4500 \pm 2.4552 $ days.
		
		In applications, these parameters must be changed to be expressed in the code-adopted units. For example, in astronomical units (years), we must use $365.25 A$ and $B+9$ instead of the given $A, B$.
		
		The variation of the stellar angular velocity corresponding to Eq. \ref{eq:logtGyr} for stars with a rotation period up to 170 days is
		\begin{equation}
			\dot{\Omega}_*= \frac{-\Omega_*^2}{2\pi At\ln(10) } %% 10^{AP_*+B}},   
	\end{equation}
	which must be added to the tidal component of $\dot\Omega_*$ in the system differential equations. 
	
	\section{Relaxation factors and time lags}\label{Q}
	
	The analysis of modern versions of Darwin's theory shows that the Quality Factor $Q$ used to characterize the response of the body to the tidal stress is not defined in the same way following whether the body's rotation is synchronized or not. When the rotation is synchronized, $Q \simeq 1/\varepsilon_2$ where $\varepsilon_2$ is the lag corresponding to the tidal component whose period is the orbital period, while otherwise, $Q$ is the inverse of the semi-diurnal lag $\varepsilon_0$
	(\cite{FRH} Sec. 11.3). 
	
	As a consequence, when the equations of Darwinian theory are compared to their equivalents in the creep-tide theory, we have to consider separately the two cases. 
	If the body rotation is synchronized with the orbital motion,
	\begin{equation}
		\varepsilon_2 = \frac{1}{Q} \simeq \left(\frac{n}{\gamma} + \frac{\gamma}{n}\right)^{-1}.
	\end{equation}
	Otherwise \cite{rheo2013}
	\begin{equation}
		\varepsilon_0 = \frac{1}{Q} \simeq \left(\frac{\nu}{\gamma} + \frac{\gamma}{\nu}\right)^{-1}.
	\end{equation}
	We note that the behavior of these functions inverts at the value corresponding to the minimum possible value $Q=2$. The constant time lag (CTL) version of Darwin's theory corresponds to the branch $ \gamma > n$, while the relaxation factors used in the creep-tide theory applied to LP 791-18d are such that $ \gamma << n$.

	\section{Used parameters} 
	The masses, radii, and orbital elements used in the calculations are those given by Peterson et al. \cite{LP791}. The other parameters used are given in table \ref{tab1}.
	
	\begin{table}[h]
		\caption{Other stellar and planetary parameters adopted}\label{tab1}
		\begin{tabular}{l@{\qquad}c@{\qquad}c}
			\toprule
			& Star  &  Planet  \\
			\midrule
			Relaxation factor (s$^{-1}$)   & $8-40$  \qquad &  $10^{-8}-10^{-6}$   \\
			Moment of Inertia ($\times mR^2$) \qquad   & 0.07   & 0.33 \\
			Fluid Love number    & 0.26   & 1.28   \\
			Rotation Period (days)       &  $> 5$  &  $-$\\
			\botrule
		\end{tabular}
	\end{table}
	
	\section{Used tidal evolution equations}
	
	The effects due to the tides on the star and on the planet LP 791-18d were calculated using the creep tide evolution equations for differentiated bodies with aligned layers \cite{hybrid}:
		\begin{eqnarray}
		\langle\dot{a}\rangle &=& -\frac{k_f n  R_N^2\overline{\epsilon}_\rho}{15a} 
		\sum_{k\in \mathbb{Z}}
		\left( 3(k-2) E_{2,k}^2\sin{2\sigma_{k}}
		+ k E_{0,k}^2 \sin{2\sigma''_{k}}\right)	
	\end{eqnarray}
	\begin{eqnarray}
		\langle \dot{e} \rangle &=&  -\frac{k_fnR_N^2\overline{\epsilon}_\rho}{30a^2 e } (1-e^2)	\nonumber\\ &&  \times \sum_{k\in \mathbb{Z}} \Big[
		3\Big(\frac{2}{\sqrt{1-e^2}}+(k-2)\Big)E_{2,k}^2\sin{2\sigma_{k}} 
		+ k E_{0,k}^2 \sin{2\sigma''_{k}} \Big].
		\label{eq:dot_e}
	\end{eqnarray}
	\begin{eqnarray}
		\langle \dot{\Omega} \rangle = - \frac{GMmR_N^2k_f\overline{\epsilon}_\rho}{5Ca^3}
		\sum_{k\in \mathbb{Z}} E_{2,k}^2  
		\sin{2\sigma_{k}}.
	\end{eqnarray}
	where $m$ is the mass of the body whose tide is being calculated and $M$ is the external body whose attraction creates the tidal bulge; $R_N$ is the radius of the outer layer of the body; $k_f$ is the fluid Love number; $\overline{\epsilon}_\rho$ is the mean flattening of the equivalent Jeans homogeneous spheroid; $C$ is the polar moment of inertia; $a$,$e$,$n$ are orbital parameters; and $E_{2,k}$ are known functions of the eccentricities (Cayley coefficients).
	
	The phases $\sigma_{k}$ and $\sigma''_{k}$ are defined by
	\begin{equation}
		\sin{\sigma_{k}}  = \frac{\nu+kn}{\sqrt{\gamma^2+(\nu+kn)^2}}; \qquad
		\sin{\sigma''_{k}} = \frac{kn}{\sqrt{\gamma^2+(kn)^2}};
	\end{equation}
	where	 $ \nu=2(\Omega-n) $ is the semi-diurnal frequency of the body and $ \gamma$ is the relaxation factor.

\end{appendices}

\bmhead{Acknowledgements}
We thank Prof. Cristian Beaugé and the referee for their careful reading of the manuscript.
This research was supported by CNPq-Brasil (Proc. 303540/2020-6), FAPESP (Procs. 2016/13750-6 and 2023/03060-6 ref. PLATO mission) and the Munich Institute for Astro-, Particle and BioPhysics (MIAPbP), which is funded by the Deutsche Forschungsgemeinschaft (DFG) under Germany´s Excellence Strategy – EXC-2094 – 390783311.

Data Availability Statement: No Data associated in the manuscript.



\begin{thebibliography}{}
	\bibitem{Rory}
	Barnes, R.: {Tidal locking of habitable exoplanets}
	Celest. Mech. Dyn. Astron. 129, 509-536 (2017)
	doi: 10.1007/s10569-017-9783-7
	
	\bibitem{Bouvier}
	Bouvier, J., Forestini, M. and Allain, S.: {The angular momentum evolution of low-mass stars}, Astron. Astrophys. 326, 1023?1043 (1997)
	ADS bibcode: 1997A\&A...326.1023B
	
\bibitem{Calleg}
Callegari, N. and Rodr\'{\i}guez, A., {Dynamics of rotation of super-Earths}. Celest. Mech. Dyn. Astron. 116, 389-416 (2013)
doi: 10.1007/s10569-013-9496-5

	\bibitem{Celletti}
	Celletti, A.: \textit{Stability and chaos in celestial mechanics}, Springer, Heidelberg, 2010.
	ISBN 978-3-540-85145-5
	
	\bibitem{Correia}
	Correia, A.C.M., Bou\'e, G., Laskar, J, and  Rodr\'{\i}guez, A.: Deformation and tidal evolution of close-in planets and satellites using a Maxwell viscoelastic rheology, Astron. Astrophys. 571, A50 (2014)
	doi: 10.1051/0004-6361/201424211
	
	\bibitem{Guinan}
	Engle, S.G. and Guinan, E.F., Living with a Red Dwarf: The Rotation-Age Relationships of M Dwarfs. Astrophys. J. Letters, 954, L50 (2023)
	doi: 10.3847/2041-8213/acf472
	
	\bibitem{DGS}
	Ferraz-Mello, S.: \textit{Dynamics of the Galilean Satellites}, IAG-USP, S\~ao Paulo (1979; revised 2022)
	ADS bibcode: 1979dgsa.book
	
	\bibitem{FRH}
	Ferraz-Mello, S., Rodr\'{\i}guez, A. and Hussmann, H., {Tidal friction in close-in satellites and exoplanets: The Darwin theory re-visited.}  Celest. Mech. Dyn. Astron.  101, 171-201 (2008) and 104, 319-320 (2009) (Errata) 
	doi: 10.1007/s10569-008-9133-x
	
	\bibitem{rheo2013}
	Ferraz-Mello, S.: {Tidal synchronization of close-in satellites and exoplanets. A rheophysical approach.} Celest. Mech. Dyn. Astron. 116, 109-140 (2013). 
	doi: 10.1007/s10569-013-9482-y
	
	\bibitem{rheo2015}
	Ferraz-Mello, S.: {Tidal synchronization of close-in satellites and exoplanets. II. Spin dynamics and extension to Mercury and exoplanet host stars.} Celest. Mech. Dyn. Astron. 122, 359-389 (2015). 
	doi: 10.1007/s10569-015-9624-5
	
	\bibitem{hybrid}
	Ferraz-Mello, S., Folonier, H.A. and Gomes, G.O., {Creep tide theory: equations for differentiated bodies with aligned layers}, Celest. Mech. Dynam. Astron. 134:
	25 (2022) 
	doi: 10.1007/s10569-022-10082-8
	
	\bibitem{rheo2018}
	Folonier, H.A., Ferraz-Mello, S., and Andrade-Ines, E.: Tidal synchronization of close-in satellites and exoplanets.
	III. Tidal dissipation revisited and application to Enceladus. Celest. Mech. Dyn. Astron. 130:78 (2018). 
	doi: 10.1007/s10569-018-9872-2
	
	\bibitem{3D}
	Folonier, H. A., Ferraz-Mello, S.,  and Alves-Silva, R., {Extension of the creep tide theory to exoplanets systems with high stellar obliquity. The dynamic tide of CoRot-3b}, Celest. Mech. Dynam. Astron. 137:15 (2025)
	doi: 10.1007/s10569-025-10245-3
	
	\bibitem{Corot7}
	L\'eger, A., Grasset, O., Fegley B., Codron, F., Albarede A.F. et al.:
	The extreme physical properties of the CoRoT-7b super-Earth,
	Icarus 213, 1–11 (2011)
	
	\bibitem{LP791}
	Peterson, M. S., Benneke, B., Collins, K., Piaulet, C., Crossfield, I. J. M. {\it et al.}, {A temperate Earth-sized planet with tidal heating transiting an M6 star}, Nature,  {617}, 701-705  (2023)
	doi: 10.1038/s41586-023-05934-8
	
	\bibitem{Adrian}
	Rodr\'{\i}guez, A., Callegari Jr., N., Michtchenko  T. A. and Hussmann, H.:
	Spin-orbit coupling for tidally evolving super-Earths, 
	Mon. Not. R. Astron. Soc. 427, 2239-2250 (2012) doi:10.1111/j.1365-2966.2012.22084.x
	
	\bibitem{Zieba}
	Zieba, S., Kreidberg, L., Ducrot, E., Gillon, M., Morley, C. et al.: No thick carbon dioxide atmosphere on the rocky exoplanet TRAPPIST-1 c. Nature, 620, 746-749 (2023) doi:10.1038/s41586-023-06232-z
	
	%\bibitem{Ref}
	
	% etc
\end{thebibliography}
\end{document}